\documentclass[aps,prl,reprint,longbibliography,superscriptaddress,floatfix]{revtex4-2}
\usepackage{graphicx}
\graphicspath{{figures/}}
\usepackage[export]{adjustbox}
\usepackage{amsmath,amssymb}
\usepackage{mathtools}
\usepackage{bm}
\usepackage{multirow}
\usepackage{bigints}
\usepackage[usenames,dvipsnames]{xcolor}
\usepackage{tabularx}
\usepackage{enumitem}
\usepackage[%
  colorlinks=true,
  urlcolor=blue,
  linkcolor=blue,
  citecolor=blue
]{hyperref}
\usepackage{mathrsfs}

\def\XXint#1#2#3{{\setbox0=\hbox{$#1{#2#3}{\int}$}
     \vcenter{\hbox{$#2#3$}}\kern-.5\wd0}}

\newcommand{\beq}{\begin{equation}}
\newcommand{\eeq}{\end{equation}}

\providecommand\bnabla{\boldsymbol{\nabla}}
\providecommand\bcdot{\boldsymbol{\cdot}}

\usepackage{xcolor} 
\definecolor{newcolor}{rgb}{.8,.349,.1}

\definecolor{lightblue}{rgb}{0.63, 0.74, 0.78}
\definecolor{seagreen}{rgb}{0.18, 0.42, 0.41}
\definecolor{orange}{rgb}{0.85, 0.55, 0.13}
\definecolor{silver}{rgb}{0.69, 0.67, 0.66}
\definecolor{rust}{rgb}{0.72, 0.26, 0.06}
\definecolor{joshua}{RGB}{251,220,127}

\colorlet{lightsilver}{silver!30!white}
\colorlet{darkorange}{orange!85!black}
\colorlet{darksilver}{silver!85!black}
\colorlet{darklightblue}{lightblue!85!black}
\colorlet{darkrust}{rust!85!black}

\begin{document}


\title{Self-organized dynamics of a viscous drop with interfacial nematic activity}


\author{Mohammadhossein Firouznia}
\affiliation{Department of Mechanical and Aerospace Engineering, University of California San Diego, \\ 9500 Gilman Drive, La Jolla, California, 92093, USA}
\affiliation{Center for Computational Biology, Flatiron Institute, New York, New York, 10010, USA\\
\vspace{-0.25cm}$\,$}
\author{David Saintillan}
\email[]{dstn@ucsd.edu}
\affiliation{Department of Mechanical and Aerospace Engineering, University of California San Diego, \\ 9500 Gilman Drive, La Jolla, California, 92093, USA}



\begin{abstract}
We study emergent dynamics in a viscous drop subject to interfacial nematic activity.\ Using hydrodynamic simulations, we show how the interplay of nematodynamics, activity-driven flows and surface deformations gives rise to a sequence of self-organized behaviors of increasing complexity, from periodic braiding motions of topological defects to chaotic defect dynamics and active turbulence, along with spontaneous shape changes and translation.\ Our findings recapitulate qualitative features of experiments and shed light on the mechanisms underpinning morphological dynamics in active interfaces.
\end{abstract}

\maketitle


Living materials are characterized by their ability to continuously transform chemical energy into mechanical work at the microscale.\ When confined to deformable surfaces, these active materials demonstrate a myriad of dynamic behaviors. Biological surfaces also often exhibit intrinsic degrees of freedom that correspond to in-plane order (such as nematic or polar), which facilitate long-range hydrodynamic interactions, resulting in the emergence of self-organized spatiotemporal patterns. This is a fundamental feature of various biological systems, including the cell cortex and confluent eukaryotic cells, and plays a crucial role in their functional properties. Here, we focus on systems with nematic symmetry.\ Experimental observations indicate the emergence of nematic order during cytokinesis \cite{spira2017cytokinesis, reymann2016cortical}.\ Nematic alignment has also been evidenced in different stages of tissue morphogenesis when individual cells exhibit a preferred elongation axis \cite{guillamat2022integer, saw2017topological, duclos2017topological}.

Biological processes from subcellular to multicellular scales often occur on self-deforming surfaces with various mechanical properties.\ Examples include shape changes during polarization, migration and division in cells, and apical constrictions in epithelial morphogenesis \cite{salbreux2012Trendscellbiology_actin, martin2014apical}. Understanding these morphological dynamics necessitates theoretical models that account for the three-way coupling between surface deformations, in-plane order, and flow \cite{mietke2019self, salbreux2022PRRes_theory,al2023morphodynamics}. Recent studies have probed the mechanisms driving shape changes in cells and epithelial tissues, advancing our understanding of these processes \cite{metselaar2019topology, ruske_Yoemans2021PRX_morphology, khoromskaia2023active}.

The central motif of these biological systems has also been utilized to create synthetic soft active materials \cite{keber2014topology, guillamat2018active, weirich2019self-org_spindle_sythetic}.\ In pioneering work, Keber \textit{et al.}~\cite{keber2014topology} assembled a shape-shifting lipid vesicle by encapsulating a film of microtubules and kinesin motors to its inner surface. Understanding the interplay between orientational order, activity-induced flow, and substrate geometry has been the subject of several studies thereafter.\ Various models have probed the role of surface curvature on the dynamics of topological defects in active fluids confined to rigid surfaces of various topologies \cite{Zhang_Depablu2016NatureComm, shankar2017topological, green2017geometry, nitschke2018nematic, PearceGiomi2019PRL_active_turbulence}, yet the role of interfacial deformations and their coupling to bulk flow remains poorly understood. 


In this Letter, we report on the spontaneous dynamics of a viscous drop driven out of equilibrium due to interfacial nematic activity. We show that the interplay between the flow inside and outside the drop, surface transport of the nematic field and surface deformations, gives rise to a sequence of self-organized behaviors and symmetry-breaking phenomena of increasing complexity. Our results recapitulate the qualitative features of experiments \cite{keber2014topology}, both in the small and finite deformation regimes. Under small deformations, the dynamics is characterized by the braiding motion of topological defects around the drop, giving rise to well-known braiding patterns at different activity levels. The asymmetry induced under finite deformations results in translational motion of the drop. Under strong activity, a transition to active turbulence is also observed.

\begin{figure*}
\centering
\includegraphics[width=0.98\textwidth]{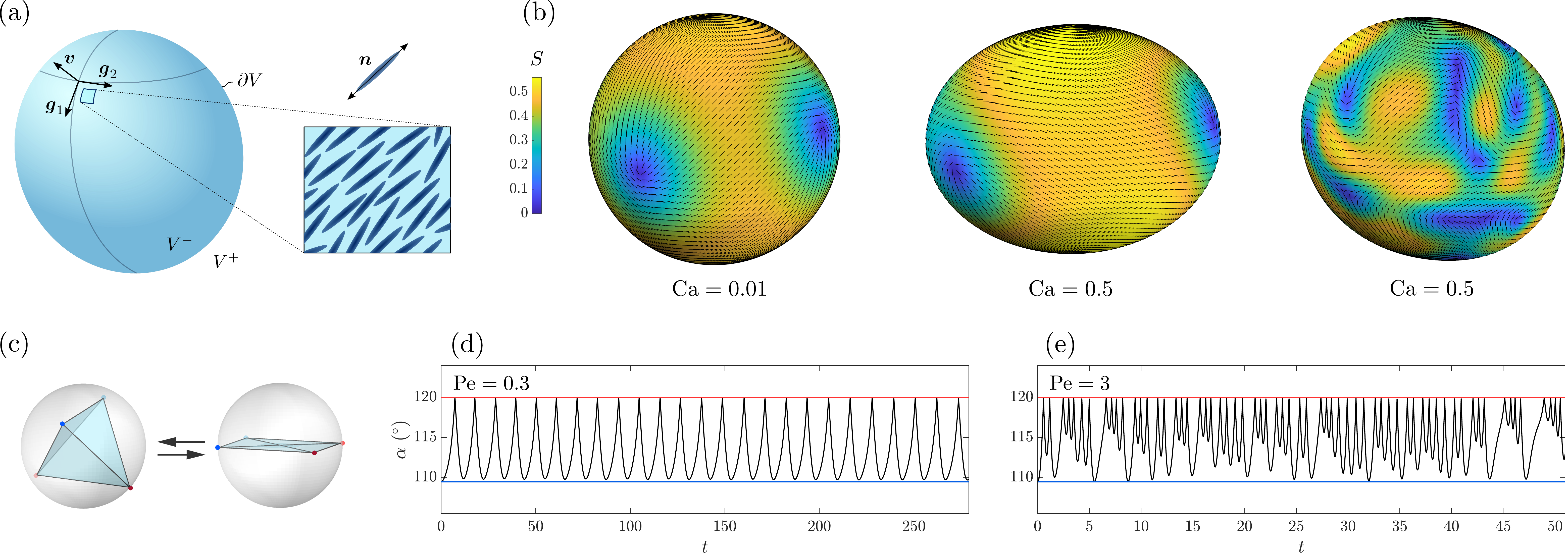} 
	\caption{(a) Schematic of the model system. (b) Snapshots of the director field and scalar order parameter shown as a colormap during: (left) braiding motion under small deformations, (middle) braiding motion with shape-shifting behavior, and (right) chaotic regime with finite deformations.\ Also see Videos 1--3 of the SM \cite{Supplemental_Info}.\ (c) Tetrahedral (left) and planar (right) defect arrangements. (d,e) Evolution of the average angle $\alpha$ for two systems with  $\mathrm{Pe}=0.3$ (d) and  $\mathrm{Pe}=3$ (e). The red and blue lines denote planar ($120^\circ$) and tetrahedral ($109.5^\circ$) configurations, respectively. $(\mathrm{Ca},\,\ell_c)=(0.01,\,0.32)$ in (d,e). \label{figs:tetrahedral_planar_oscillations}} 
\end{figure*}

We model a viscous drop occupying volume $V^-$ and suspended in another viscous fluid $V^+$ [Fig.~\ref{figs:tetrahedral_planar_oscillations}(a)].\ The interface $\partial V$ is defined by a smooth surface $\bm{r}=\bm{r}(s^1,s^2)\in\mathbb{R}^3$ parameterized by coordinates $(s^1,s^2)$.\ Tangent vectors $\bm{g}_i=\partial_i \bm{r}$ ($\partial_i \! \coloneqq\partial/\partial s^i$, $i=1,2$), along with the unit normal $\smash{ \bm{v}={(\bm{g}_1\times \bm{g}_2)}/{|\bm{g}_1\times \bm{g}_2|} }$, form a local coordinate system with surface metric tensor $g_{ij}=\bm{g}_i \bcdot \bm{g}_j$. The drop is initially spherical with radius $R$.\ A monolayer of active nematic particles is constrained to $\partial V$ and drives the system out of equilibrium by inducing flows inside and outside the drop and by causing deformations. 
We use a coarse-grained representation of the nematic field in terms of the tensor $\bm{Q}=Q^{ij}\,\bm{g}_i\bm{g}_j$ where $Q^{ij}=S(n^i n^j-g^{ij}/2)$. Here, $\bm{n}=n^{i}\bm{g}_i$ ($n^i n_i=1$) denotes the local nematic director, and $S\in[0,1]$ is the scalar order parameter characterizing the strength of alignment.\ The $\bm{Q}$ tensor evolves by the nematodynamic equation \cite{giomi2015PRX, PearceGiomi2019PRL_active_turbulence, salbreux2022PRRes_theory}
\begin{align}
 \mathrm{D}_t Q^{ij} = \dfrac{1}{\Gamma} H^{ij} + \zeta \widetilde{U}^{ij}, \qquad \boldsymbol{x}\in \partial V.\label{eq:nematodynamics}
\end{align}
$\mathrm{D}_t Q^{ij}$ is a co-rotational material derivative embodying advection and rotation of the nematic by the surface flow with velocity $\bm{u}=u^i\bm{g}_i +u_n \bm{v}$: 
\begin{equation}\label{eq:DtQ_corot_def, eq:DtQ_corot_def}  
\begin{split}
 \mathrm{D}_t Q^{ij}= &~\partial_t Q^{ij} + u^k \, \nabla_k Q^{ij} +u_n \,(C^{j}_{k}Q^{ik} + C^{i}_{k}Q^{kj}) \\
 &\qquad + \omega_n \,(\epsilon^{ik} Q^{j}_{k}+ \epsilon^{jk} Q^{i}_{k}),
\end{split}
\end{equation}
where $\nabla_k$ is the covariant derivative with respect to $s^k$, $C_{ij}=-\partial_j \partial_i \bm{x} \bcdot  \bm{n}$ is the curvature tensor,  $\omega_n=\frac{1}{2}\epsilon^{ij}\nabla_i u_j$ is the normal vorticity, and $\epsilon^{ij}=\bm{v}\bcdot (\bm{g}^i\times \bm{g}^j) $ is the Levi-Civita tensor.\ The molecular tensor $H^{ij}=-\delta \mathcal{F}/ \delta Q^{ij}$ describes orientational relaxation in the nematic monolayer, with $\Gamma$ the rotational viscosity. It derives from the Landau--de Gennes free energy,
\begin{equation}\label{eq:free_energy_def}
 \mathcal{F}\!=\!\!\int_{\partial V} \!\!\!\!\mathrm{d}A \Bigl(  \frac{k_s}{2}\!\left[d\, Q_{ij}Q^{ij} \!+\!c(Q_{ij}Q^{ij})^2\right]
 \!+\! \frac{k_e}{2} \nabla_i Q_{jk} \nabla^i Q^{jk}\! \Bigr),
\end{equation}
which accounts for short-range and elastic interactions in the nematic monolayer \cite{Voigt2020liquid, kralj2011curvature}.\ Here, $k_s$ and $k_e$ are steric and elastic phenomenological constants, and $d=a+\frac{b}{3}S+\frac{c}{6}S^2$ where $a$, $b$ and $c$ are normalized thermotropic parameters. We assume that the characteristic size of a nematic particle is small compared to the local radius of curvature of the interface, and thus neglect any coupling between the extrinsic curvature and the nematic tensor in Eq.~\eqref{eq:free_energy_def}.  Finally, alignment by the flow is captured by  $\smash{\zeta \widetilde{U}^{ij}}$ where $\zeta$ is the flow alignment parameter or Bretherton constant \cite{bretherton1962}, and $\widetilde{U}^{ij}$ is the traceless strain-rate tensor:
\beq
\widetilde{U}^{ij}={U}^{ij}-\tfrac{1}{2}U^k_k g^{ij}, \,\,\,  {U}^{ij}=\tfrac{1}{2}(\nabla^i u^j+\nabla^j u^i)+ C^{ij} u_n. 
\label{eq:strain_rate_tensor_def}
\eeq

Neglecting inertial effects and gravity, the flow inside and outside the drop is governed by the Stokes equations:
\begin{equation}
{\mu^{\pm}}\nabla^2\bm{u}^{\pm}-\bnabla p^{\pm}=\mathbf{0},    \quad   \bnabla \bcdot\bm{u}^{\pm}=0,   \quad  ~\bm{x}\in V^{\pm}. \label{eq:stokes_continuity}
\end{equation}
The velocity is continuous across $\partial V$ and vanishes far away. The nematic particles exert an active surface stress $\bm{T}^a=\xi \, \bm{Q}$ on their surrounding, leading to fluid motion and deformations.  The constant $\xi$ captures the biochemical activity, with $\xi<0$ for extensile systems as in the experiments of Keber \textit{et al.}~\cite{keber2014topology}. The local force balance along the tangential and normal directions  on the drop surface $\partial V$ reads
\begin{eqnarray}
& f^{h,j}  + \xi \, \nabla_i Q^{ij}=0, &   \label{eq:force_bal_tang}\\
	& {f^h_n  -\gamma \,C_k^k-\,\xi \, C_{ij}\, Q^{ij}=0, }& \label{eq:force_bal_normal} 
\end{eqnarray} 
where $\gamma$ is the uniform surface tension.\ The jump in hydrodynamic tractions across $\partial V$ is $\bm{f}^h=\bm{v}\bcdot[\bm{T}^{h}   \big] _{-}^{+}$, where $\bm{T}^{h}=-p\bm{I}+\mu\big(\bnabla\bm{u}+\bnabla\bm{u}^\mathrm{T}\big)$ is the Newtonian stress tensor.\ 
We solve Eqs.~\eqref{eq:nematodynamics}--\eqref{eq:force_bal_normal} numerically using a custom spectral boundary integral solver \cite{Firouznia2023_JCP, Firouznia_SBIM_2022_Github}, where all physical variables such as shape,  velocity, and $\boldsymbol{Q}$ tensor are represented as truncated series of spherical surface harmonics (see Supplemental Material (SM) \cite{Supplemental_Info} for details).

Dimensional analysis of the governing equations yields four dimensionless groups in addition to $\zeta$.\ The active capillary number $\mathrm{Ca}={|\xi|\, R}/{\gamma}$ compares the strength of active stresses to surface tension and governs the magnitude of deviations from the spherical shape [Fig.~\ref{figs:tetrahedral_planar_oscillations}(b)]. The active P\'eclet number $\mathrm{Pe}={|\xi| \, \Gamma}/{(k_s \mu^+ R^2)}$ characterizes the strength of convective vs relaxational fluxes in Eq.~\eqref{eq:nematodynamics}. The balance between short-range and elastic interactions in the nematic monolayer defines a dimensionless coherence length $\ell_c=R^{-1}\sqrt{k_e/ k_s}$, or effective distance over which topological defects affect the nematic field on the scale of the drop. Finally, the viscosity ratio between the inner and outer fluids is $\lambda=\mu^-/\mu^+$.\ We found that varying $\lambda$ has little effect on the dynamics, and all results shown here are for  $\lambda=1$.

\begin{figure*}
\centering
\includegraphics[width=0.99\textwidth]
{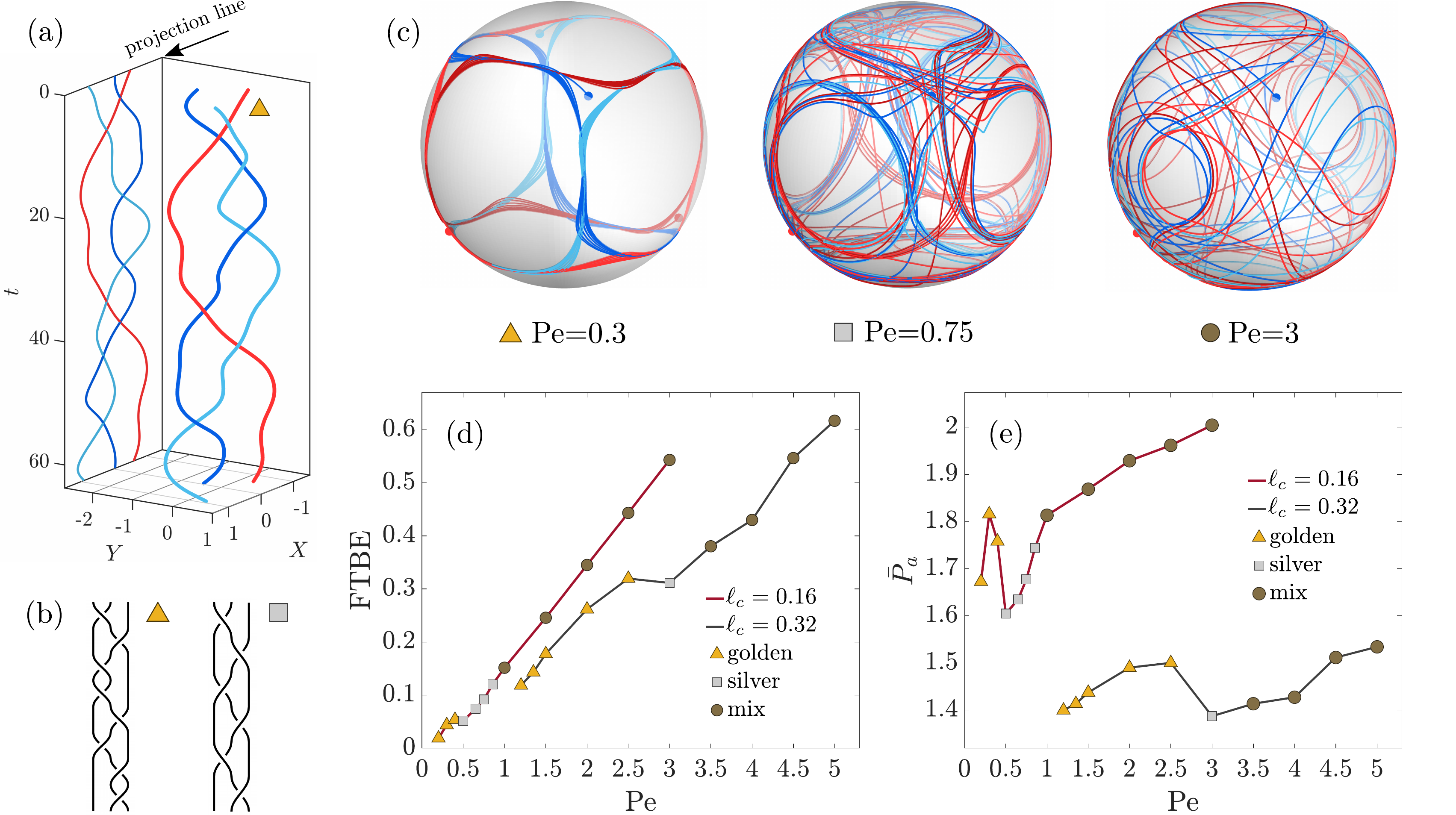}\vspace{0.15cm}
	\caption{(a) Spatiotemporal trajectories $(X,Y,t)$ of projected defects for a drop with $(\mathrm{Pe},\ell_c) = (0.3,0.16)$ during golden braiding pattern (see SM for details \cite{Supplemental_Info}). (b) Diagrams of golden (triangle) and silver (square) braids. (c) Defect trajectories in 3D for active drops with $\ell_c=0.16$ in the golden (left), silver (middle), and mix (right) braiding regimes at different values of $\mathrm{Pe}$.  The time interval between the first and last data points is $\Delta t=50/f_1$, where $f_1$ corresponds to the peak frequency in the FFT spectrum of $\alpha$. (d) FTBE of defect trajectories, and (e) average active power $\bar{P}_a$ for a spherical active drop, as functions of $\mathrm{Pe}$ and $\ell_c$. The markers identify golden, silver, and mix braiding regimes.   $\mathrm{Ca}=0.01$ in all cases.}\label{figs:FTBE_power_3Dtrajectories_Ca1e-2}  
\end{figure*}

First, we focus on the regime of $\mathrm{Ca}\ll1$ to isolate the effect of deformations.\ In this regime, the behavior of the system is governed by two parameters:\ the activity level captured by $\mathrm{Pe}$, and nematic elasticity captured by $\ell_c$. Active stresses drive the system out of equilibrium while nematic elasticity tends to stabilizes it. Indeed, as nematic elasticity becomes stronger, the coherence length increases ($\ell^2_c\propto k_e$), resulting in a more constrained arrangement of the topological defects that repel one another. Given the spherical topology of the interface, the net topological charge of the nematic field is fixed at $+2$, and for small to moderate activity levels ($0<\mathrm{Pe}\lesssim5$), the system exhibits four $+1/2$ defects [Fig.~\ref{figs:tetrahedral_planar_oscillations}(b), left]. 

Under small activity levels, the defects remain stationary and occupy the vertices of a regular tetrahedron (Fig.~S2 \cite{Supplemental_Info}).\ Beyond a critical value $\mathrm{Pe}_c$, a transition occurs to an unsteady regime marked by the periodic motion of the defects (see Video 1 \cite{Supplemental_Info}).\ Each defect is set into motion by the activity-induced flow, which influences the nematic field.\ In agreement with experimental observations \cite{keber2014topology}, the defect arrangement oscillates between tetrahedral and planar configurations, as shown in Fig.~\ref{figs:tetrahedral_planar_oscillations}(c-e), where $\alpha=1/6\sum_{i<j}^{4} \alpha_{ij}$ denotes the average of all pairwise angles $\alpha_{ij}$ between the four defects.



The spatiotemporal defect trajectories can be analyzed and interpreted using the concept of {braids} \cite{thiffeault2022braids}.\ We first map the defect trajectories from the spherical surface of the drop to its two-dimensional mid-plane using the stereographic projection, where one reference defect is taken to be at the sphere pole and therefore mapped to infinity \cite{fadell1962braid, birman1974braids}. When projected along a specific direction, the remaining three defect trajectories display a sequence of crossings that defines a braid $\bm{b}$ [Fig.~\ref{figs:FTBE_power_3Dtrajectories_Ca1e-2}(a)]. Each crossing event is expressed in terms of elementary generators $\sigma_i,~i\in\{1,\dots, n-1\},$ of the $n$-particle braid group with $n=3$ \cite{thiffeault2022braids}.\ The generator $\sigma_i$ denotes the clockwise exchange of defect $i$ with defect $i+1$ on the projection line, while $\sigma_{i}^{-1}$ corresponds to their counterclockwise exchange. The complexity of a given braid and its mixing efficiency can be quantified by the exponential stretching rate of material curves in the associated flow.\ This is determined through the braid's topological entropy and finite-time braiding exponent (FTBE) for periodic and aperiodic trajectories, respectively \cite{boyland2000topological, fathi1979travaux, FTBE_budivsic2015finite} (see SM for details \cite{Supplemental_Info}).


For activity levels close to $\mathrm{Pe}_c$ within the unsteady regime, the defect motion is periodic and characterized by a single time scale, as evidenced by the presence of a single peak in the fast Fourier transform (FFT) of the average angle $\alpha$ (Fig.~S4 \cite{Supplemental_Info}).\ The defect trajectories follow a specific pattern known as the \textit{golden braid} [Fig.~\ref{figs:FTBE_power_3Dtrajectories_Ca1e-2}(a,b)], described by the braid $\bm{B}_1={(\bm{b}_1)}^k$, where $\bm{b}_1=\sigma_1 \, \sigma_{2}^{-1} \, \sigma_1 \, \sigma_{1} \, \sigma_2 \, \sigma_{1}^{-1}\, \sigma_{2}^{-1} \, \sigma_{2}^{-1}$.\ Its topological entropy is given by $h(\bm{b}_1)\!=6\log{\phi_1}$, where $\phi_1=(1+\sqrt{5})/2$ denotes the golden ratio \cite{Alessandro_goldenbraid1999mixing}.\ As $\mathrm{Pe}$ is further increased, nonlinear effects become more pronounced, leading to the emergence of a second time scale.\ This is manifested in the FFT spectrum of $\alpha$, which exhibits two distinct peaks (Fig.~S5 \cite{Supplemental_Info}).\ Eventually, the braiding pattern undergoes a transition from the golden braid to the \textit{silver braid} $\bm{B}_2={(\bm{b}_2)}^k$, where $\bm{b}_2=\sigma_1 \, \sigma_{2} \, \sigma_{1}^{-1} \, \sigma_{2}^{-1} \, \sigma_1 \, \sigma_{2}^{-1}$.\ Its topological entropy is $h(\bm{b}_2)=2\log{\phi_2}$, where  $\phi_2=1+\sqrt{2}$ is the silver ratio \cite{finn2011topological}.\ Schematic diagrams of the golden and silver braids are shown in Fig.~\ref{figs:FTBE_power_3Dtrajectories_Ca1e-2}(b).\ The efficiency of a periodic braid in increasing entropy can be quantified by its topological entropy by generator (TEPG) \cite{finn2011topological}.\ Comparing the TEPG of the golden and silver braids shows that $\mathrm{TEPG}(\bm{b}_1)\approx 1.23\mathrm{TEPG}(\bm{b}_2)$, i.e., the transition to the silver braid with increasing $\mathrm{Pe}$ causes a decrease in TEPG.


As the activity level is further increased, a transition from periodic to aperiodic dynamics is observed.\ This regime, referred to as \textit{mix braiding}, is characterized by the absence of a specific braiding pattern, with defect trajectories displaying irregular and chaotic behavior.\ The FFT spectrum of $\alpha$ in this regime is broad and displays multiple peaks, a signature of highly nonlinear dynamics (Fig.~S6 \cite{Supplemental_Info}).\ Figure \ref{figs:FTBE_power_3Dtrajectories_Ca1e-2}(c) shows 3D defect trajectories in drops with golden, silver, and mix braiding patterns, providing visual evidence of the increasing complexity with higher activity. To characterize the different braiding patterns and quantify their complexities in the non-periodic case, we calculate the FTBE of defect trajectories as a function of $\mathrm{Pe}$ and $\ell_c$ in Fig.~\ref{figs:FTBE_power_3Dtrajectories_Ca1e-2}(d) \cite{FTBE_budivsic2015finite, braidlab}. The FTBE generally shows an increasing trend with respect to $\mathrm{Pe}$, indicating that higher activity levels lead to more complex dynamics. However, the transition from golden to silver braiding is accompanied by a decrease in FTBE, consistent with the decrease in TEPG discussed above.\ This reduction in the FTBE is particularly intriguing as it occurs despite the heightened activity. 

 
To elucidate this effect, we analyze the drop's energetics in the unsteady regime.\ 
The active power expended by the surface nematic is defined as $\smash{{P}_a(t)=\int_{_{\partial V}} \bm{f}^a\bcdot \bm{u} \,\mathrm{d}s}$, where $\smash{\bm{f}^a=\xi \nabla_i Q^{ij} \bm{g}_j-\,\xi C_{ij}\, Q^{ij} \bm{v}}$ is the interfacial active traction and $\bm{u}$ the interfacial velocity.\ As shown in Fig.~\ref{figs:FTBE_power_3Dtrajectories_Ca1e-2}(e), the time-averaged active power, denoted by $\bar{P}_a=\langle P_a \rangle_{t}$, is an ascending function of activity except during the transition from the golden to silver braid, where it suddenly drops.\ This indicates that the increase in TEPG with $\mathrm{Pe}$ in the golden braiding regime comes at a higher energetic cost, and suggests that the spontaneous transition to the silver braid is energetically favorable.\
  Note that, under small deformations, the active power is dissipated primarily through viscous effects within the bulk, with negligible contributions from capillarity (see SM for details \cite{Supplemental_Info}).   

We highlight the stabilizing effect of nematic elasticity as captured by $\ell_c$.\ The critical P\'eclet number $\mathrm{Pe}_c$ for the transition from equilibrium to periodic braiding rises from $\mathrm{Pe}_c \approx 0.2$  to $1.2$ as the coherence length is varied from $\ell_c=0.16$ to $0.32$.\ The onset of aperiodic defect motions (mix braiding regime) is also delayed under larger coherence lengths. According to Fig.~\ref{figs:FTBE_power_3Dtrajectories_Ca1e-2}(d,e), for a given $\mathrm{Pe}$, both the FTBE and the active power  are consistently lower at the larger coherence length.\ Indeed, the leading effect of nematic elasticity is to repel topological defects, which fosters more organized and constrained defect dynamics and delays the transition to chaotic motion. 


Allowing for finite deformations ($\mathrm{Ca}>0$) further increases the complexity of the dynamics. In this regime, the active drop undergoes spontaneous shape changes in the form of breathing motions as the defects traverse its surface [Fig.~\ref{figs:tetrahedral_planar_oscillations}(b), middle] (See Video 2 \cite{Supplemental_Info}). These deformations, in turn, impact the nematodynamics, establishing a three-way feedback loop between shape, nematic field, and flow. Notably, we observe an increase in $\mathrm{Pe}_c$ at higher capillary numbers, indicating that elevated activity levels are required for defects to overcome the energy barriers induced by deformations.

For the smaller coherence length $\ell_c=0.16$, we observe that under moderate values of $\mathrm{Pe}>\mathrm{Pe}_c$, four defects exhibit braiding motion similar to the behavior observed in the small deformation limit. Concurrently, the drop undergoes spontaneous shape changes under the influence of active stresses.\ At long times, we find that the drop eventually reaches an equilibrium state with a steady deformed shape and defect configuration.\ This behavior is only observed up to a second critical P\'eclet number $\mathrm{Pe}_{eq}>\mathrm{Pe}_c$, beyond which the system transitions to a chaotic regime characterized by the rapid creation and annihilation of defects around the drop [Fig.~\ref{figs:tetrahedral_planar_oscillations}(b), right] (also see Video 3 and Fig.~S9 \cite{Supplemental_Info}).\ The newly formed defects emerge as pairs with $\pm1/2$ topological charges, maintaining a constant net topological charge. The dynamics in this regime resembles the active turbulence previously observed in nematic materials on flat and curved surfaces \cite{guillamat2017taming,  ellis2018NaturePhys_curvature, lin2021energetics, PearceGiomi2019PRL_active_turbulence, doostmohammadi2018active}. For the larger coherence length $\ell_c=0.32$, the system reaches equilibrium over long times for all explored Péclet numbers ($0<\mathrm{Pe}<12$) due to the stabilizing effect of nematic elasticity.


\begin{figure}[t]	
\centering
\includegraphics[width=0.44\textwidth]{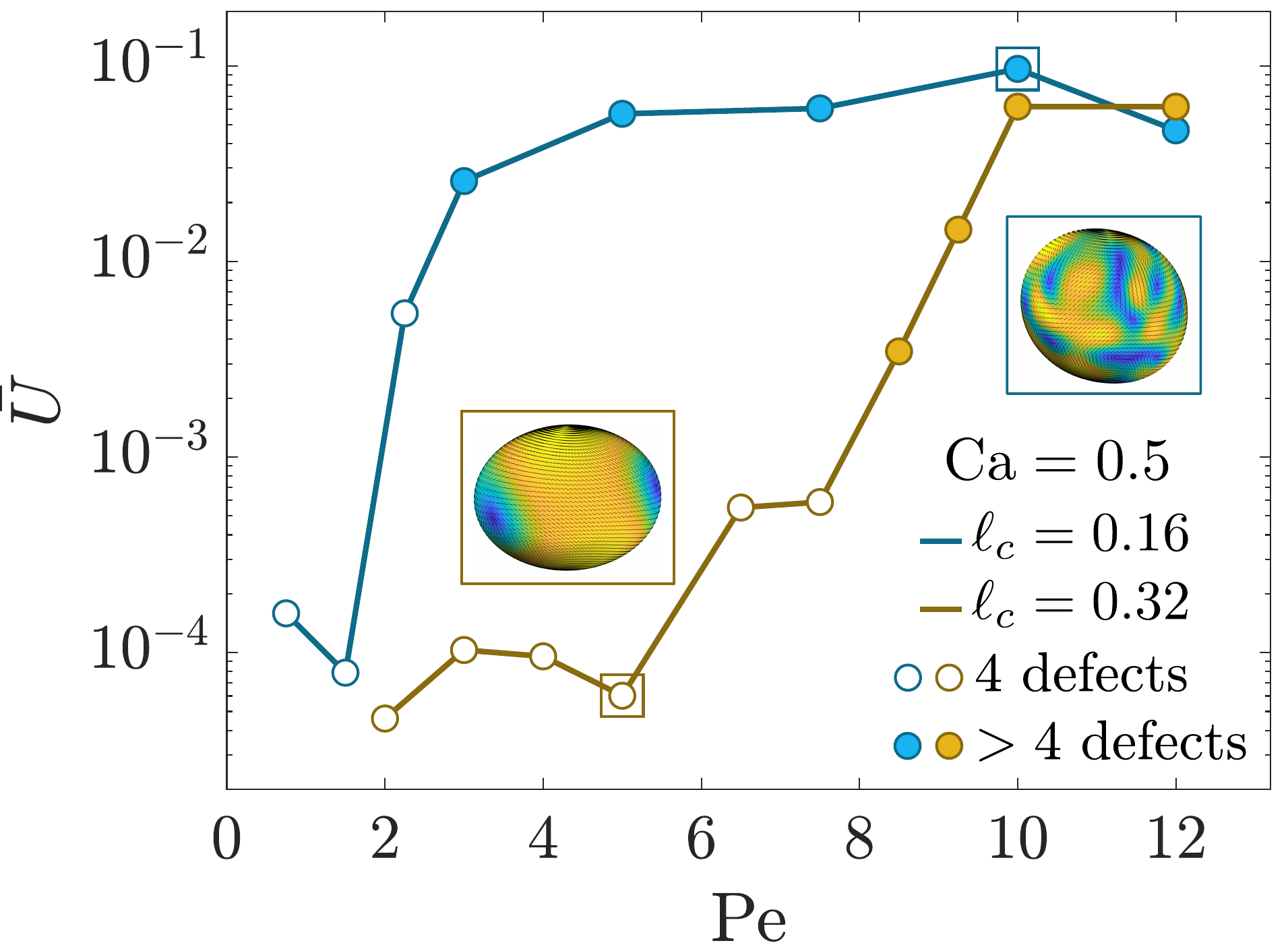}	
	\caption{Time-averaged translational velocity as a function of $\mathrm{Pe}$ for $\mathrm{Ca}=0.5$.\ Blue and brown lines correspond to $\ell_c=0.16$ and $0.32$, respectively.\ Filled markers indicate cases where more than four defects were observed. Insets show snapshots of the drop corresponding to the boxed markers. }
	\label{figs:finite_deformation_TransVel}  
\end{figure}

Nonlinearities introduced by finite deformations amplify the asymmetry in the nematic field and defect configuration.\ This asymmetry gives rise to a net translational component in the velocity field.\ The volume-averaged velocity of the drop of volume $V_d$ is calculated as $\smash{\bm{U}(t)=\frac{1}{V_{d}}\int_{{V^-}}\!\bm{u} \, \mathrm{d}\upsilon }$, and we denote by $\bar{U}=|\langle \bm{U} \rangle_t|$ the magnitude of its time average.  We observe that the translational velocity is orders of magnitude larger compared to the regime of $\mathrm{Ca}\ll1$, underscoring the role of deformations in breaking the system's symmetry.\ Figure \ref{figs:finite_deformation_TransVel} shows $\bar{U}$ vs $\mathrm{Pe}$ for $\mathrm{Ca}=0.5$.\ For $\ell_c=0.16$, $\bar{U}$ first exhibits a sharp increase with $\mathrm{Pe}$ when four defects are present. It then reaches a plateau as $\mathrm{Pe}$ is further increased and the drops enters the chaotic regime where additional defects are created.\ 
Similar trends are observed for $\ell_c=0.32$, although the increase in $\bar{U}$ with $\mathrm{Pe}$ is delayed due to the stabilizing effect of nematic elasticity, which promotes more symmetric shapes. 

Using numerical simulations, we have analyzed the dynamics of a deformable viscous drop subject to interfacial nematic activity.\ Our results highlight the complex interplay of nematodynamics, active fluid flows, and interfacial mechanics, and point to a wide range of emergent dynamics depending on the importance of active stresses relative to viscous and capillary stresses.\ In the low deformation regime, the dynamics is characterized by four +1/2 defects whose trajectories can be described in terms of braids of increasing complexity.\ At finite capillary numbers, breathing deformations and translational motion emerge, as well as an active turbulent regime with continuous generation and annihilation of defect pairs. Our observations are all consistent with past experiments on active vesicles \cite{keber2014topology}. We note that the analysis of braiding motions only provides qualitative information on the system’s mixing efficiency, and is limited to regimes with four topological defects, where treating defects as material points is a reasonable approximation.\ A more in-depth characterization of the organizing role of activity-induced flows may rely on an analysis of Lagrangian coherent structures and associated finite-time Lyapunov exponents, which have recently been applied to identify flow attractors and repellers in bulk active nematics \cite{serra2023defect, sinigaglia2023optimal} as well as during embryogenesis \cite{serra2020dynamic}. Finally, we note that the system studied here may serve as a simplified model for active living systems, such as cells or organoids.\ For that purpose, various model extensions may be desirable, including accounting for the role of elastic stresses, of extrinsic curvature coupling, or of chemical cues, which are instrumental in regulating the cell cortex \cite{lessey2012mechanical, bement2015activator, bischof2017cdk1}.

\begin{acknowledgments}
	The authors thank J.-L. Thiffeault for insightful conversations on braids and topological entropy, and S. H. Bryngelson for his contributions to the computational framework \cite{Firouznia_SBIM_2022_Github}.\ This work was partially funded by National Science Foundation Grants CBET-1934199 and DMS-2153520.

\end{acknowledgments}


\bibliography{Refs_ActiveDrop}

\end{document}